%% file: main.tex
\documentclass[a4paper, 10pt, conference]{IEEEtran}

\usepackage{algpseudocode}
\usepackage{amsfonts}
\usepackage{url}

\usepackage{graphicx}

\usepackage{epsf} 
\usepackage{booktabs}
\usepackage{array}
\usepackage[font=normalsize]{caption}
\usepackage{verbatim}

\usepackage{amsmath}
\usepackage{amsthm}
\usepackage[numbers, sort]{natbib}

\usepackage[ruled,linesnumbered]{algorithm2e}

\usepackage{titlesec}

\def\BibTeX{{\rm B\kern-.05em{\sc i\kern-.025em b}\kern-.08em T\kern-.1667em\lower.7ex\hbox{E}\kern-.125em}}
\usepackage[left=1.62cm,right=1.62cm,top=1.9cm, bottom=4.4cm]{geometry}
\begin{document}

\title{Design and Evaluation of an NDN-Based Network for Distributed Digital Twins}

\author{\IEEEauthorblockN{Chen Chen\IEEEauthorrefmark{1}, Zihan Jia\IEEEauthorrefmark{2}, Ze Wang\IEEEauthorrefmark{2}, Lin Cui\IEEEauthorrefmark{3}, Fung Po Tso\IEEEauthorrefmark{2}
  		}\\
	\IEEEauthorblockA{
        \IEEEauthorrefmark{1}Department of Computer Science and Technology, University of Cambridge, UK\\
        \IEEEauthorrefmark{2}Department of Computer Science, Loughborough University, UK\\
         \IEEEauthorrefmark{2}Department of Computer Science, Jinan University, Guangzhou, China\\
		Email: cc2181@cam.ac.uk; z.jia@lboro.ac.uk; z.wang6@lboro.ac.uk; tcuilin@jnu.edu.cn; p.tso@lboro.ac.uk
	}
    }


\maketitle

\input{abstract.tex}

	\pagestyle{empty}  
	\thispagestyle{empty} 
\thispagestyle{empty}
\input{intro}
\input{relatedwork}

\input{preliminaries}
\input{solution}

\input{exp}
\input{conclusion}

\bibliographystyle{unsrt}
{\footnotesize
\bibliography{reference}}

\end{document}

%% file: abstract.tex
\begin{abstract}
Digital twins (DT) have received significant attention due to their numerous benefits, such as real-time data analytics and cost reduction in production. DT serves as a fundamental component of many applications, encompassing smart manufacturing, intelligent vehicles, and smart cities.
By using Machine Learning (ML) and Artificial Intelligence (AI) techniques, DTs can efficiently facilitate decision-making and productivity by simulating the status and changes of a physical entity.
To handle the massive amount of data brought by DTs, it is challenging to achieve low response latency for data fetching over existing IP-based networks.
IP-based networks use host addresses for end-to-end communication, making data distribution between DTs inefficient.
Thus, we propose to use DTs in a distributed manner over Named Data Networking (NDN) networks.
NDN is data-centric where data is routed based on content names, dynamically adjusting paths to optimize latency.
Popular data is cached in network nodes, reducing data transmission and network congestion.
Since data is fetched by content names, users and mobile devices can move freely without IP address reassignment.
By using \textit{in-network caching} and \textit{adaptive routing}, we reckon NDN is an ideal fit for Future G Networks in the context of Digital Twins.
We compared DTs in edge scenarios with cloud scenarios over NDN and IP-based networks to validate our insights.
Extensive simulation results show that using DT in the edge reduces response latency by 10.2$\times$.
This \textit{position} paper represents an initial investigation into the gap in distributed DTs over NDN, serving as an early-stage study.
\end{abstract}

%% file: intro.tex
\section{Introduction}\label{sec::intro}

Digital Twins (\textbf{DT}) create a virtual copy of elements and services of a physical entity~\cite{YSUJOI20}. 
DTs leverage Machine Learning (ML) and Artificial Intelligence (AI) to monitor real-world data, process information, and generate valuable insights.
Nowadays, DTs are widely used in industries such as manufacturing and maintenance, improving the efficiency and productivity of the manufacturing process~\cite{MSCDD22}.
These advancements pave the way for integration between real networks and data analytics to facilitate timely decisions.
Particularly, Digital Twins use a data-driven model to reflect the relationships between the physical and virtual instances in a two-way manner, making data findable, accessible and interoperable.

Adopting DT brings numerous benefits to existing cloud services such as smart grids, smart cities, surveillance, smart manufacturing, image recognition and many others~\cite{chen2024, liang2023, chen2023}.
For many applications, DTs need to represent the digital copy of the real instance to receive raw data and provide services.
Examples include complicated systems like buildings, factories or vehicles.
In this context, DTs can deliver a large volume of data with low latency to a traditional IP network. Eventually, the data arrives at a cloud-based DT model.
Meanwhile, the advances in AI and ML technologies also contribute to the sheer volume of data in DT services.
As a result, two different types of data movement need to be supported: (1) \textit{physical-virtual communications}, i.e., data flow between physical instance and its digital twin; (2) \textit{inter-DT communications}, i.e., data flow between different DTs for federation purpose \cite{GTEK24}.
By this means, we envision that future DTs will be federated entities that communicate and share data with each other to make decisions.

However, transmitting a large amount of data between digital twins will massively burden existing IP-based networks.
The latency of the data stream is essential because a low quality of data stream will result in performance degradation in DTs.
Existing DTs, such as AWS IoT TwinMaker~\cite{awsdt} and Azure Digital Twins~\cite{Azuredt}, provide a cloud-based solution where data are transmitted to a centralized endpoint via IP networks, resulting in data network bottlenecks and increased latency for DTs.
This limitation becomes even more pronounced when dealing with complex systems like aircraft, power plants, or cities, involving multiple DTs and extensive data collection from physical entities.
Significant delays in data retrieval and processing can degrade DT performance, as incomplete data hinders decision-making.
For example, if the data of one sensor is not completely delivered, it will preclude DTs from making decisions.

To address the complexity of data management brought by DT networks, we propose to use Named Data Networking (NDN).
Our insight is that NDN is self-organizing, self-optimizing and self-healing, paving the way for data-driven networks.
NDN, as a data-centric network paradigm, enables consumers to fetch data by an \textit{Interest Packet} with prefixes, enabling data retrieval from multiple sources dynamically.
NDN differs from conventional IP-based networks in three aspects: (1) \textit{Pending Interest Table} keeps track of outstanding Interest packets that have not been satisfied.
(2) \textit{Forward Strategy} sends the interest packet toward the next hop based on the Forwarding Information Base (FIB).
(3) \textit{Content Store} performs as a temporary cache in NDN routers that stores data packets to optimize data retrieval and reduce redundant requests.
Using NDN networks contributes to data-driven networks in the following aspects: (1) NDN allows nodes to store and cache data, reducing redundant transmissions.
(2) NDN is self-organizing by routing data based on \textit{Interest} names rather than IP addresses, making it data-centric and more efficient.
(3) NDN supports efficient data distribution with built-in caching, enabling data reuse, reducing congestion and improving response latency.

The purpose of this work is twofold: (1) evaluate the performance of data delivery in NDN and IP-based networks for DTs, showing that NDN is a good fit for DT solutions.
(2) evaluate the performance of Edge-DT and Cloud-DT, showing the benefits of distributing DT to edge networks.
Edge-DT refers to using DTs in a few edge nodes where Cloud-DT refers to aggregating all DTs in a cloud data center.
DTs require real-time synchronization with their physical counterparts.
Distributed DTs allow computation to happen where the data resides, shortening the latency in data communication.
Also, many applications, such as healthcare and finance, disallow data sharing due to privacy regulations.
Distributed DT can use federated learning to keep data local, improving privacy and security.

Thus, we compared different scenarios in this work, i.e., cloud DT in NDN, edge DT in NDN, cloud DT in IP networks and edge DT in IP networks.
We summarize our main contributions as follows.

\begin{itemize}
    \item We propose an architecture to distribute DTs in the edge over NDN networks, facilitating data management and reducing latency of fetching data.
    \item We compare cloud DT and edge DT in NDN and IP-based networks, showing that edge DT in NDN can efficiently deal with data retrieval and data caching.
    \item We implement the simulations via Network Simulator 3 (NS3), justifying the superiority of edge DT and NDN networks.
    Extensive simulation results justify that edge DT reduces response latency by 46.7\% compared to cloud DT.
    Similarly, DT in NDN networks reduces response latency by 10.2$\times$ compared to that of IP-based networks.
\end{itemize}


%% file: relatedwork.tex
\section{Related work}\label{sec::relatedwork}

\paragraph{Digital Twin over IP Networks}

To facilitate bidirectional data exchange between the physical asset and its digital twin, existing DT systems primarily employ an IP network-based model. 
Cloud platforms have become a preferred host for digital twin systems due to the need for substantial computing resources to support model learning, data analysis, and various services that enhance functionality~\cite{RSK20}.

Alves et al.~\cite{ASMTKSAL19} developed a smart farming digital twin using a cloud-based IoT agent to control an irrigation system driven by AI or the farmer.
Angin et al.~\cite{AAGGB20} presented AgriLoRa, a low-cost farmland digital twin framework for smart agriculture. It integrates a wireless sensor network for data collection and cloud-based computer vision algorithms for analyzing plant health.
Fahim et al.~\cite{FSCCD22} proposed a digital wind farm system where physical turbines connect via a 5G network, and virtual entities are hosted on Microsoft Azure Digital Twins. The physical and virtual farms communicate via a Representational State Transfer (REST) Application Programming Interface (API).
Nasiri and Kavousi-Fard~\cite{NK23} 
introduced an energy hub digital twin system on Amazon Cloud Service to enhance the electrical grid resiliency.
Issa et al.~\cite{ISA23} proposed a power prediction digital twin for a wind turbine's generic model, enabled by a real-time data streaming structure of Azure Digital Twin~\cite{Azuredt}.
Höfgen et al.~\cite{HVBZKVM23} developed a digital twin system for a robot assembly process using socket-based communication. The system is built on Azure Digital Twins.

Wang et al.~\cite{WTS23} introduced an innovative AWS cloud-based digital twin solution tailored for emergency healthcare scenarios. 
The medical equipment utilizes the Message Queuing Telemetry Transport (MQTT) protocol to forward messages to the IoT Gateway of AWS IoT Core, which is integrated into the broader AWS Cloud infrastructure.
Deshpande et al.~\cite{DSK24} proposed a 5G RAN digital twin deployed on Azure Digital Twins.
The authors investigated the performance of Azure Digital Twins in terms of latency and lag. 
Their experiments showed a linear relationship between both metrics and model/update size. 
Additionally, the digital twin in the cloud consistently experiences a long-tail latency effect.

In general, IP-based networks are host-centric rather than data-centric, falling short of managing massive amounts of data brought by DTs.

\paragraph{Digital Twin over Named Data Networking} 
Unlike IP-based deployment, only a few works have explored deploying DTs on NDN.  
Liang et al.~\cite{liang2023} proposed an NDN-based DT architecture for data management in IoT networks. 
They evaluated the architecture's packet transmission latency and drop rate under various conditions, including bandwidth, data query rate, and the number of nodes. 
The results show that both metrics decrease with increasing bandwidth but rise as the data query rate increases. 
Amadeo et al.~\cite{AMRN23} studied service discovery and provisioning problems in social DTs.
The authors proposed a centralized architecture that applies NDN to DTs.  
Compared to the legacy solution, their proposal reduced the number of hops required to retrieve data.  
Later on, Amadeo et al.~\cite{Amadeo2024} extended the work in~\cite{AMRN23} to a distributed approach.  
The service discovery latency of their solution outperformed that of two centralized approaches, which deploy services located at cloud/edge servers.  

Current NDN approaches do not consider distributed DTs and hence we are interested in investigating the performance of distributed DT over NDN.

%% file: preliminaries.tex
\section{Preliminaries}
\paragraph{Digital Twin}
Digital Twins (DTs) serve as virtual replicas of physical entities, mirroring their functionalities, components, and services in a digital environment. By leveraging Machine Learning (ML) and Artificial Intelligence (AI), DTs continuously monitor real-world data, process information, and generate insights to enhance decision-making. DTs have already been widely adopted in industries such as smart manufacturing, maintenance, and industrial automation. 

A key characteristic of DTs is their bidirectional synchronization with physical systems. Any change in the physical entity is reflected in the DT, and modifications in the DT can also influence the physical counterpart. This synchronization enables real-time system monitoring, predictive maintenance, and automated decision-making. Given the complexity of modern cyber-physical systems (CPS), DTs need to efficiently manage and process vast amounts of data to maintain low latency and high reliability.

DTs operate within a multi-layered structure, depending on the physical system they represent. This includes infrastructure-level DTs (e.g., smart grids, connected vehicles) and application-level DTs (e.g., industrial automation, healthcare monitoring). Two essential types of data exchange occur in DT systems:

Physical-to-virtual communication – Data continuously flows from the real-world entity to its DT to maintain an up-to-date digital state.
Inter-DT communication – DTs share data and collaborate with other DT instances to make federated decisions.
As DTs continue to evolve, their ability to integrate with cloud, edge, and network-based architectures becomes crucial. Existing DT implementations, such as AWS IoT TwinMaker and Azure Digital Twins, rely on cloud-based infrastructures where data is transmitted through traditional IP networks. However, the growing volume of DT-generated data poses significant challenges, including network congestion, data retrieval delays, and inefficient resource management. To address these issues, new architectures are needed to improve scalability, reliability, and responsiveness in DT systems.

\paragraph{Named Data Networking}

Named Data Networking (NDN) is a data-centric network architecture that enables content retrieval based on names rather than IP addresses. Unlike conventional IP networks, where communication is data-centric and relies on end-to-end connections, NDN operates on a data-centric paradigm, allowing users to retrieve data by requesting a specific content name rather than contacting a specific server. Existing Digital Twin (DT) systems rely heavily on IP-based networking for data transmission. However, with the increasing volume of data generated by DTs, traditional IP networks face significant challenges in bandwidth utilization, data caching, and transmission latency. As a result, NDN has emerged as a promising alternative for optimizing DT data management.

One of the key distinctions between NDN and IP-based networks lies in their data retrieval mechanisms. In IP networks, data transmission is established via host-to-host communication, requiring devices to establish a session using IP addresses. In contrast, NDN leverages a content-oriented approach, where data is requested using Interest Packets that specify the desired content name. The network then forwards the request and returns the corresponding Data Packet without the need for a fixed host location. This mechanism is particularly advantageous in DT environments, where frequent and dynamic data access is required. By employing distributed caching and intelligent forwarding strategies, NDN significantly reduces network congestion and improves data retrieval efficiency.

NDN consists of three fundamental components:

\textbf{Pending Interest Table (PIT):}
Stores outstanding Interest Packets that have not yet been satisfied, enabling efficient request management and eliminating redundant transmissions.

\textbf{Forwarding Information Base (FIB):}
Directs Interest Packets toward the appropriate next-hop based on content names rather than IP addresses, ensuring efficient data delivery.

\textbf{Content Store (CS):}
Functions as a caching mechanism, storing previously requested data packets to reduce dependence on remote servers and minimize network overhead.

Existing cloud-based DT solutions—such as AWS IoT TwinMaker and Azure Digital Twins—rely on centralized data transmission models, which can lead to network congestion and increased latency. In contrast, NDN aligns with the decentralized nature of DTs by utilizing in-network caching and optimized forwarding mechanisms, thereby reducing redundant data transmissions and improving overall data flow management.

In the context of DT systems, NDN provides several key advantages:

\textbf{Efficient Data Retrieval:} Instead of relying on fixed IP addresses, devices can request data directly using Interest Packets from different hosts, enhancing data accessibility and robustness.

\textbf{Optimized Caching Mechanism:} The Content Store (CS) in NDN routers allows frequently requested data to be cached closer to the consumer, reducing latency and improving network efficiency.

\textbf{Facilitating Data Sharing and Collaboration: }NDN enables inter-DT communication without requiring dedicated end-to-end connections, allowing multiple DTs to share data efficiently.

Given these benefits, NDN is a promising networking paradigm for managing DT data transmission, contributing to data-driven networks. In the following sections, we will further explore the application of NDN in DT environments and evaluate its performance under different deployment scenarios, including cloud-based DTs and edge-based DTs.

%% file: solution.tex
\section{NDN-assisted Distributed DT}\label{sec::solution}

In this section, we envision the proposed approach, namely, NDN-assisted distributed DT. We first demonstrate the framework and then describe the evaluation scenarios.

\begin{figure}[htbp]
    \centering
    \smallskip
     \includegraphics[width=0.3\textwidth]
    {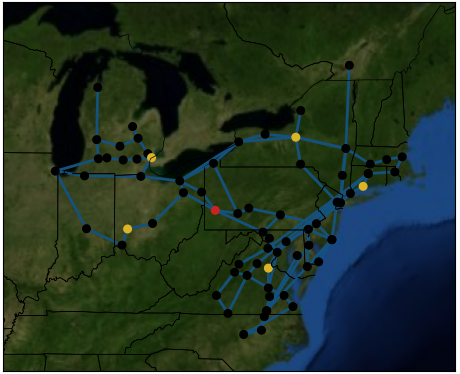}  
    \caption{Topology of Intellifiber}
    \label{fig::topo}
\end{figure}

\subsection{Framework}
\begin{figure}[htbp]
    \centering
     \includegraphics[width=0.45\textwidth]
    {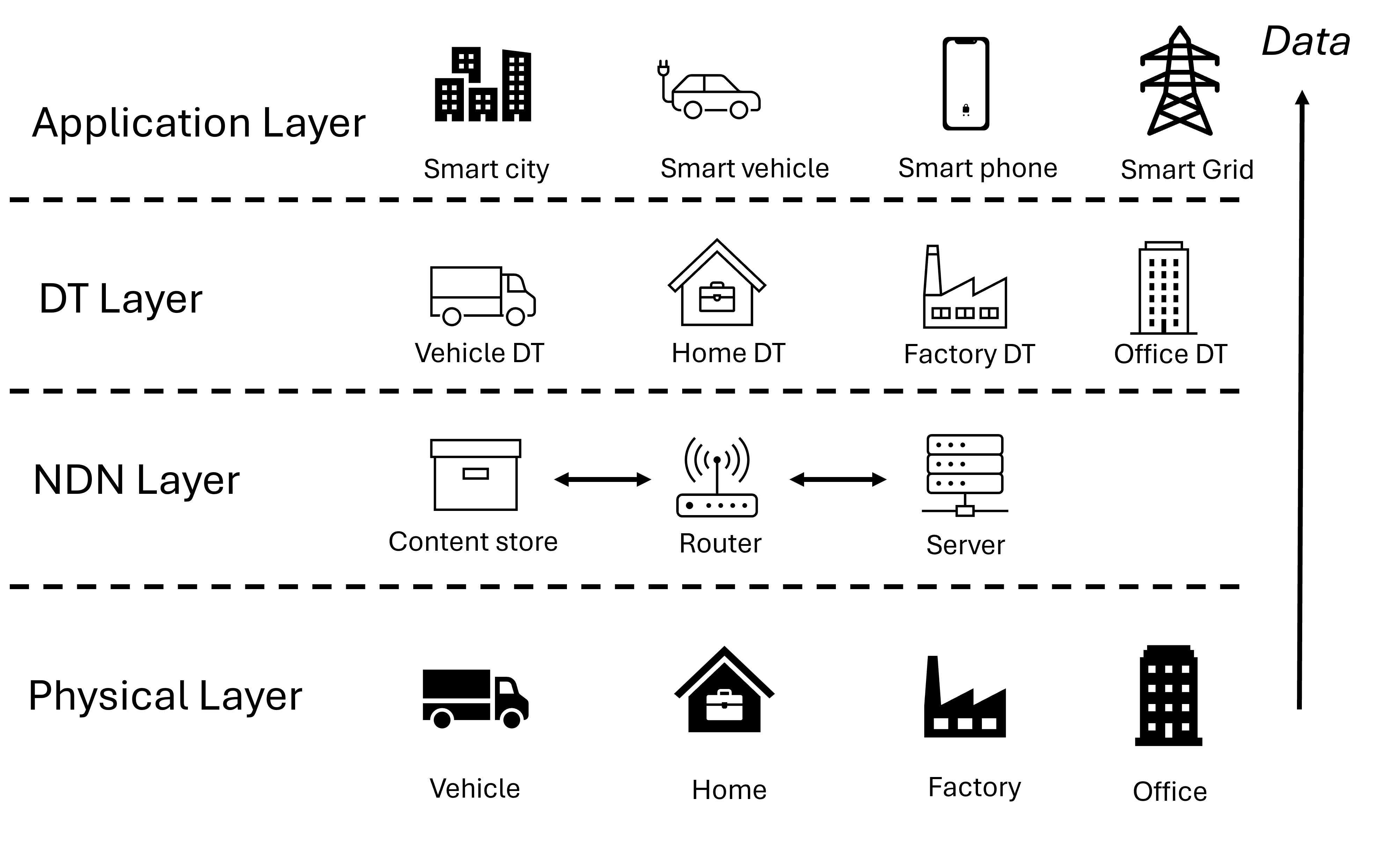}  
    \caption{NDN-assisted Digital Twins}
    \label{fig::framework}
\end{figure}

We propose a four-layer architecture for the NDN-assisted DT framework as illustrated in Figure~\ref{fig::framework}.
We have used the Intellifiber topology from the Internet Topology Zoo~\cite{topologyzoo} with 73 nodes and 97 links.
Edge DT is shown in yellow, black nodes are data consumers, red node is the set of Cloud DTs and links are shown in blue.

The proposed framework consists of four layers: (1) \textit{Physical Layer} represents the physical entities that generate raw data for the DT layer, including IoT devices, vehicles, buildings and etc.
(2) \textit{NDN Layer} is using NDN networks to collect data from the physical layer for the digital twin layer.
NDN layer uses NDN-enabled routers to cache and forward data so the end-to-end latency can be massively reduced.
Also, data is queried by using \textit{Interest packet}.
(3) \textit{DT Layer} refers to the set of digital copies of the physical entities in the physical layer.
Machine learning and data analytic models are implemented in the DT models to process the raw data and provide insightful information to the applications.
(4) \textit{Application Layer} refers to the applications that receive the decisions made by DT layer based on the raw data fetched from the physical layer.
The application layer may also send instructions back to the physical layer via the NDN layer.

\subsection{Cloud and edge-DT scenario}
\paragraph{Cloud based scenario}
The cloud-based scenario refers to deploying 5 Digital Twins in a centralized cloud (red node), shown in Figure~\ref{fig::topo}.
Consumers send queries from different locations in the network.
The cloud-based DTs have a set of DTs and the required data will be sent back to consumers via NDN or IP-based network.
For NDN networks, routers have content stores to cache the required data and may respond to consumers if specific data is available.
In IP-based networks, queries need to go through a large number of hops before reaching the host.
Moreover, certain popular links to the host may be overloaded at peak loads in IP-based networks.

\paragraph{Edge based scenario}
For the edge-based scenario, we have used 5 digital twins distributed in different locations (yellow nodes) as shown in Figure~\ref{fig::topo}.
Each of them only serves one type of query.
When a consumer starts a query, the query may be routed to a nearby Edge DT if required data is available.
By this means, communication latency is reduced by avoiding sending data to a remote cloud.
On the other hand, the distributed Edge DT will also affect the cache hit in the content store because cached data is also distributed compared to the centralized DT.

%% file: exp.tex
\section{Experimental results}\label{sec::exp}

We conduct our simulations on a server with 32GB RAM, 14 cores 13th Gen Intel(R) i7-13700H 2.40 GHz CPUs.
The server runs ndnSim 2.7~\cite{ndnSIM} over Network Simulator 3~\cite{Riley2010} version 3.29 and Ubuntu 20.04 LTS.

\subsection{Experiment setup}

We have selected 5 types of representative DTs, including \textit{Robotic Arms}, \textit{Health Care}, \textit{Vehicle Performance}, \textit{Machine Maintenance} and \textit{Asset Management}~\cite{Michael2023}.

\paragraph{Performance benchmarks:}
\begin{itemize}
    \item \textbf{\textit{NDN-Edge}} uses five digital twins in five edge nodes (the yellow node in Figure~\ref{fig::topo}) for data query over the NDN network.
    Each node only represents one type of DTs.
    \item \textbf{\textit{NDN-Cloud}} uses five digital twins in one cloud node (the red node in Figure~\ref{fig::topo}) for data query over the NDN network.
    In other words, the cloud data centre represents 5 types of DTs.
    \item  \textbf{\textit{IP-Edge}} uses five digital twins in five edge nodes (the yellow node in Figure~\ref{fig::topo}) for data query over an IP-based network.
     \item  \textbf{\textit{IP-Cloud}} uses five digital twins in one cloud node (the red node in Figure~\ref{fig::topo}) for data query over an IP-based network.
\end{itemize}

\paragraph{Topology}
As illustrated in Figure~\ref{fig::topo}, the bandwidth of links is set in [10, 200] Mbps with [0, 50] ms delay.
The content store of NDN network is set to the Least Recently Used (LRU) replacement policy.
Each consumer sends 20/60 Interest packets per second and each packet is 1024 bytes.

\subsection{Simulation with 20 Interest packets per second}

\begin{figure}[htbp]
    \centering
\includegraphics[width=0.35\textwidth]{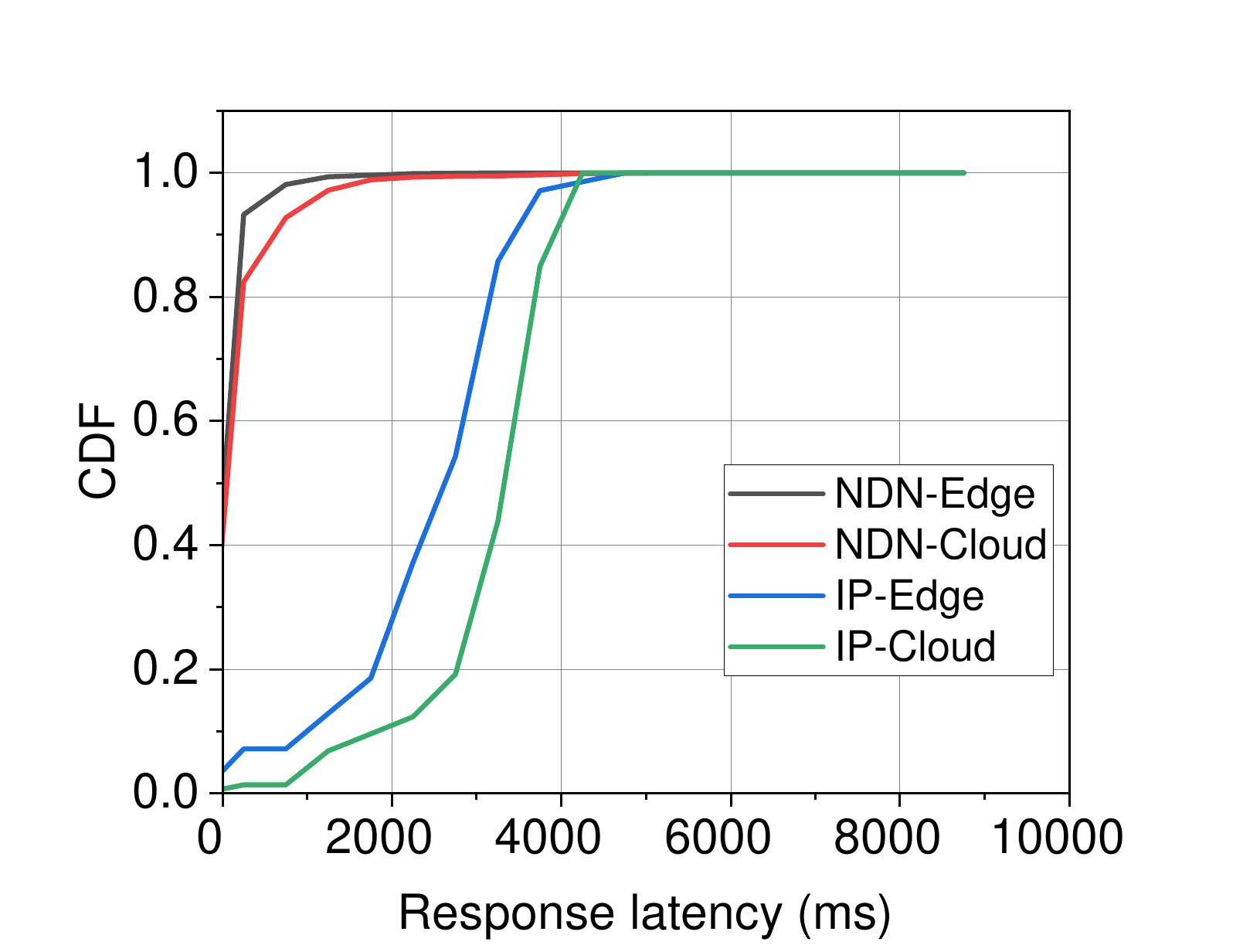}  
    \caption{CDF of response latency}
    \label{fig::latency-100k}
\end{figure}

We first examine the simulation results when consumers send 20 Interest packets per second.
Figure~\ref{fig::latency-100k} shows the Cumulative Distribution Function of response latency for all queries in the network.
It is not surprising that NDN-Edge receives the championship with P99 latency at 1250 ms followed by NDN-Cloud at 2250 ms.
In contrast, P99 latency of IP-Edge and IP-Cloud are around 4250 ms and 4500 ms, respectively.
The results first justify NDN networks have more advantages over IP-based networks due to the data caching mechanism.
Second, using distributed DT can help reduce latency because consumers can fetch data from nearby DTs.

\begin{figure}[htbp]
    \centering
\includegraphics[width=0.35\textwidth]{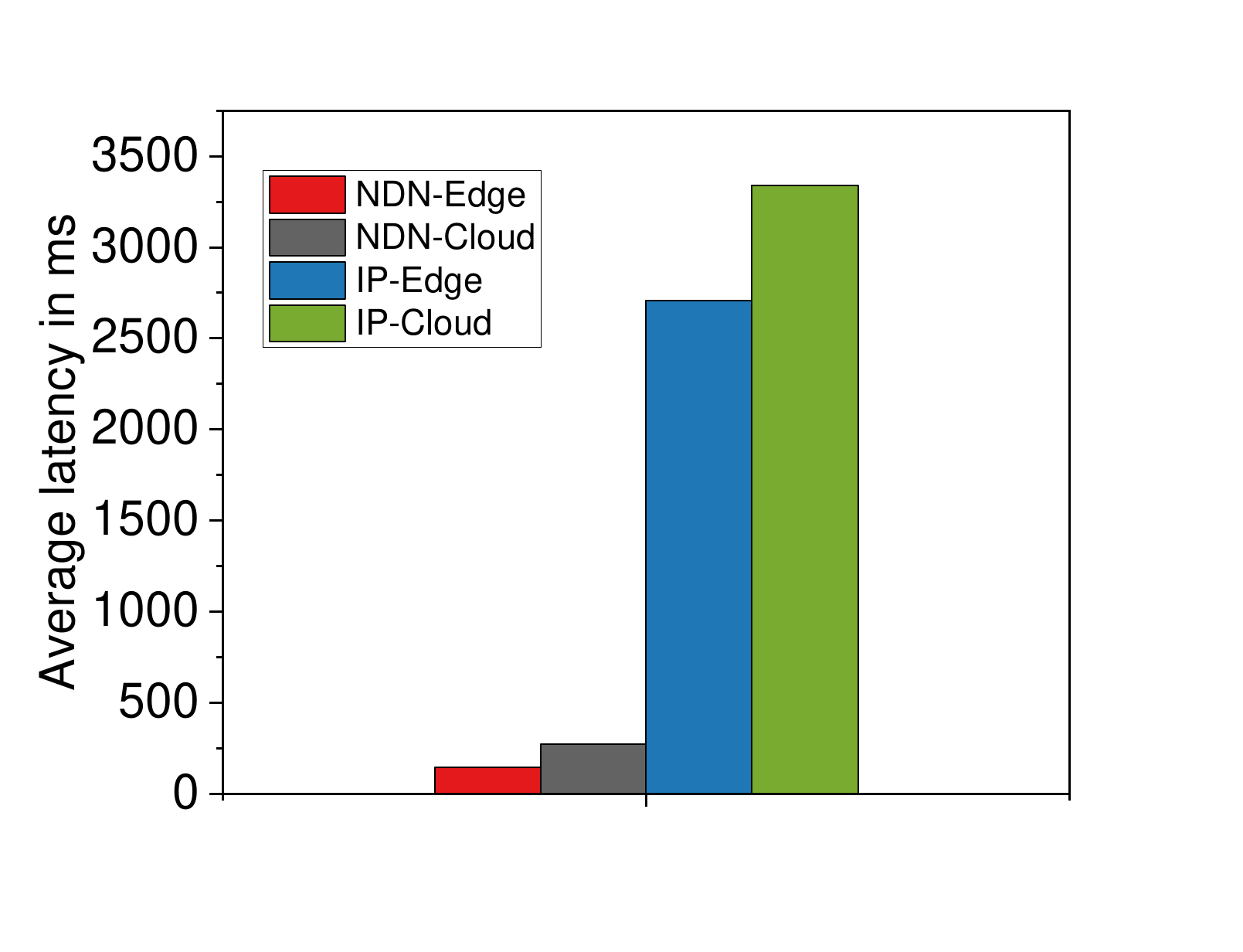}  
    \caption{Average response latency}
    \label{fig::avg-100k}
\end{figure}

Figure~\ref{fig::avg-100k} shows the average response latency achieved by different benchmarks. 
The results indicate that NDN-Edge outperforms other benchmarks.
The average response latency of NDN-Edge is around 145.9 ms while that of NDN-Cloud is 272.3 ms.
The average response latencies for IP-based networks are 2707.8 ms and 3339.6 ms, respectively.
The rationale is that NDN networks can cache data in routers and hence enable consumers to fetch data from nearby routers.
Furthermore, distributing DTs to the edge enables the data to be processed at the proximity of consumers.

\begin{figure}[htbp]
    \centering
\includegraphics[width=0.35\textwidth]{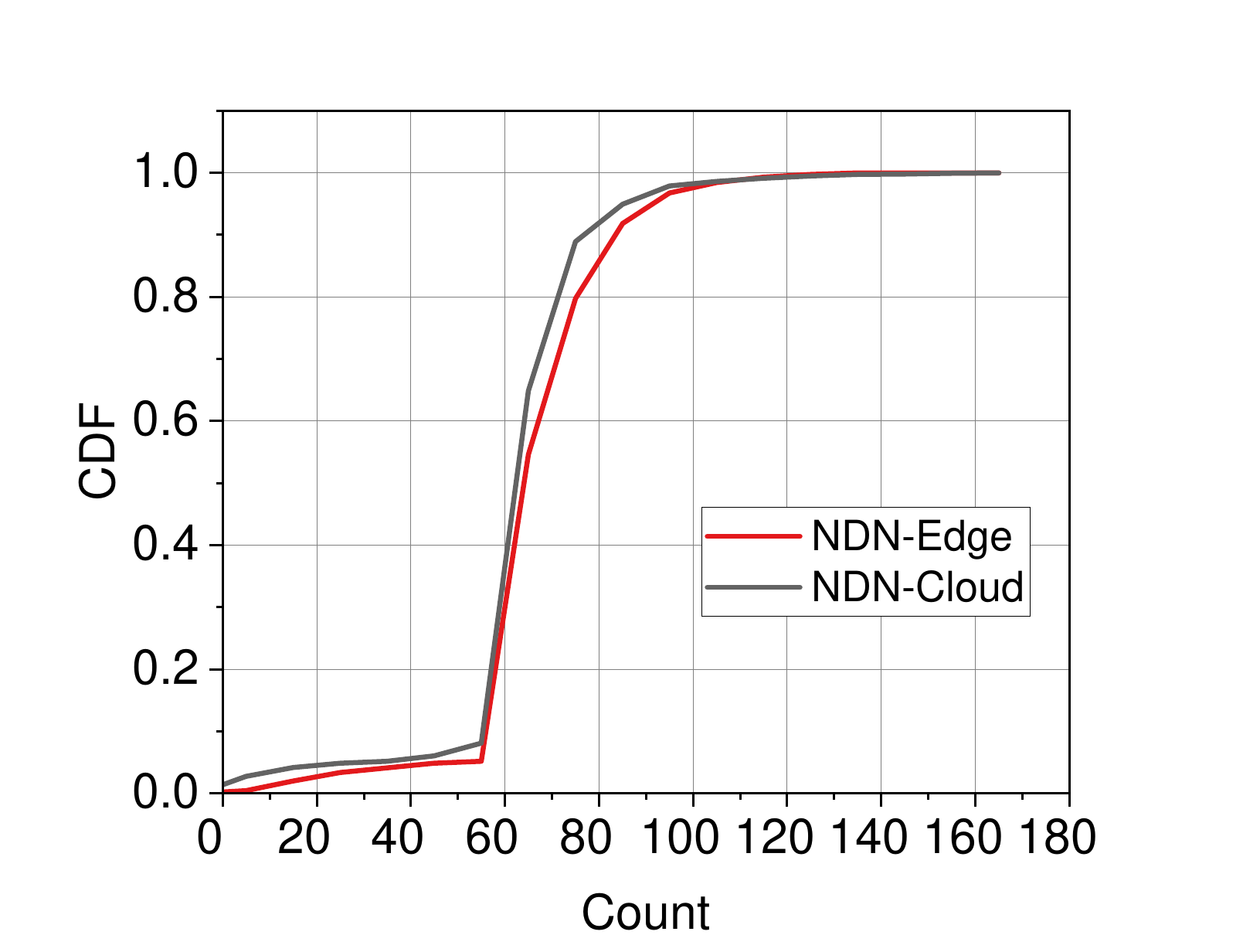}  
    \caption{Count of cache hits}
    \label{fig::cache-100k}
\end{figure}

Figure~\ref{fig::cache-100k} examines the cache hits on each node in the NDN network. 
Since IP-based networks do not have this mechanism, we cannot report its result.
As expected, the NDN-Edge approach receives the best performance in terms of \textit{CacheHits}.
Distributing DT to the edge can make data cache less skewed in popular links and thus improve cache hits.
The P95 number of cache hits for the Edge DT is around 95 while that of Cloud DT is around 85.
This indicates that Edge DT efficiently improves the cache hits and hence consumers can reuse data nearby.

\subsection{Simulation with 60 Interest packets per second}
Now we examine the simulation results when consumers send 60 Interest packets per second, justifying that the proposed approach is efficient under different workloads.

\begin{figure}[htbp]
    \centering
\includegraphics[width=0.35\textwidth]{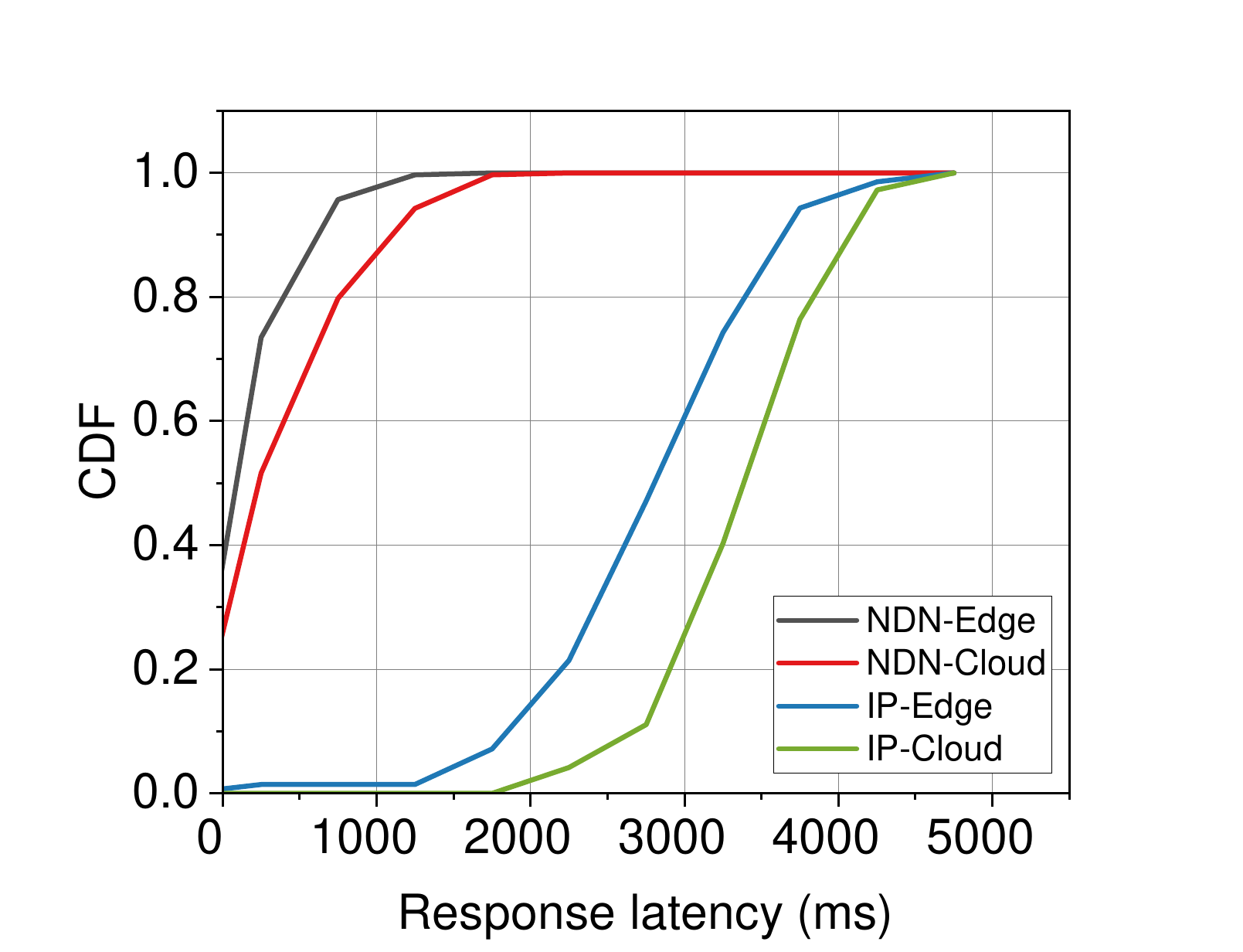}  
    \caption{CDF of response latency}
    \label{fig::100k-60rps}
\end{figure}

As shown in Figure~\ref{fig::100k-60rps}, we observe similar trends compared to Figure~\ref{fig::latency-100k}.
NDN-Edge receives the best performance, followed by NDN-Cloud.
The P99 latency of NDN-Edge and NDN-Cloud are around 1500 and 1800 ms, while that of IP-Edge and IP-Cloud are around 4500 ms.
These results justify that: (1) Running DTs in NDN networks efficiently reduces response latency for DTs. (2) Distributing DTs to the edge can further reduce the response latency.

\begin{figure}[htbp]
    \centering
\includegraphics[width=0.35\textwidth]{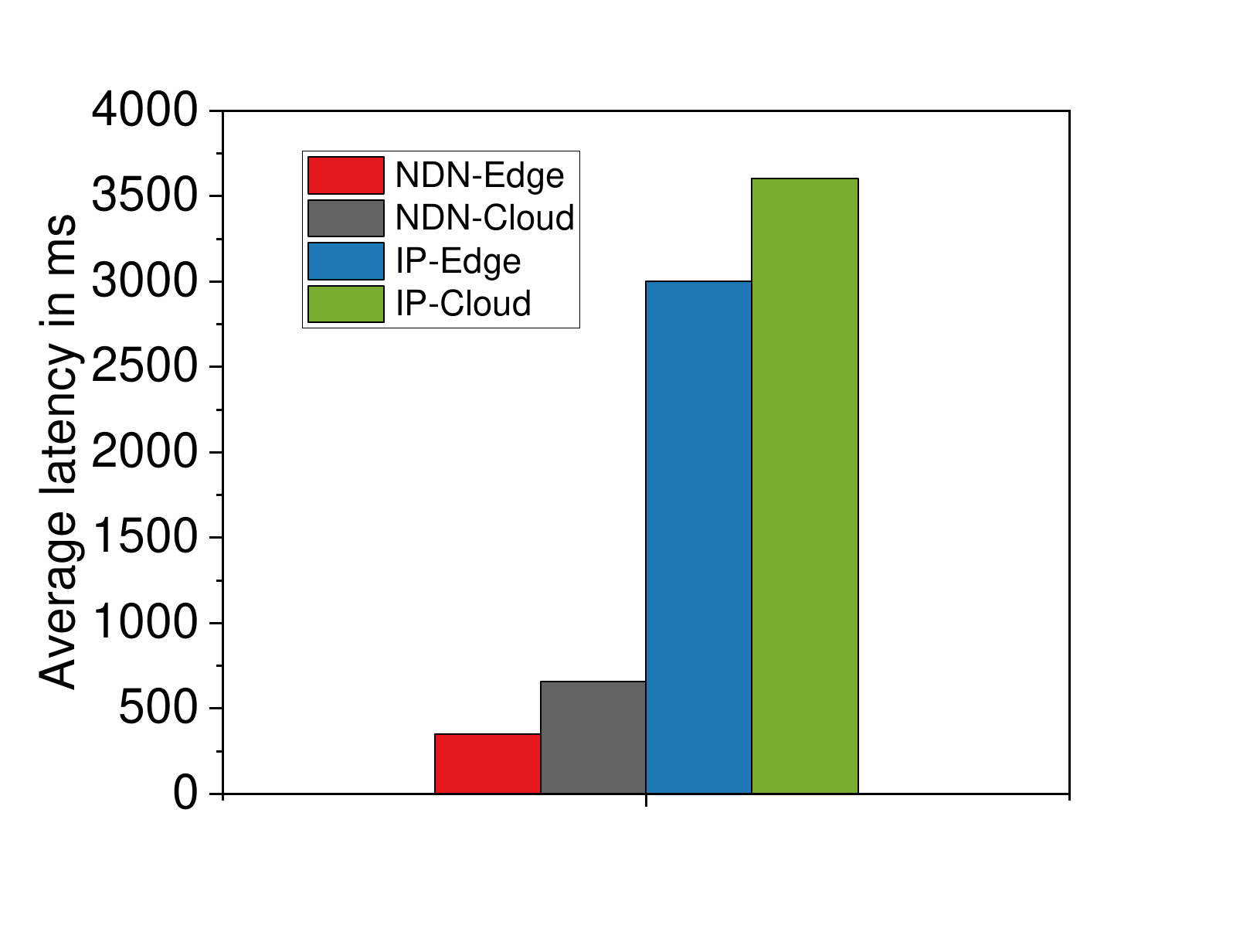}  
    \caption{Average response latency}
    \label{fig::avg-100k-60rps}
\end{figure}
Figure~\ref{fig::avg-100k-60rps} shows the average response latency of all approaches.
As expected, NDN-Edge outperforms other approaches by 53.2\% due to data caching and its distributed nature.
This result also indicates that NDN-Edge is an efficient approach for data-driven networks.
NDN allows routers to store and serve data, reducing redundancy and enhancing efficiency. This is especially useful for digital twin networks with intermittent or decentralized connectivity.

\begin{figure}[htbp]
    \centering
\includegraphics[width=0.35\textwidth]{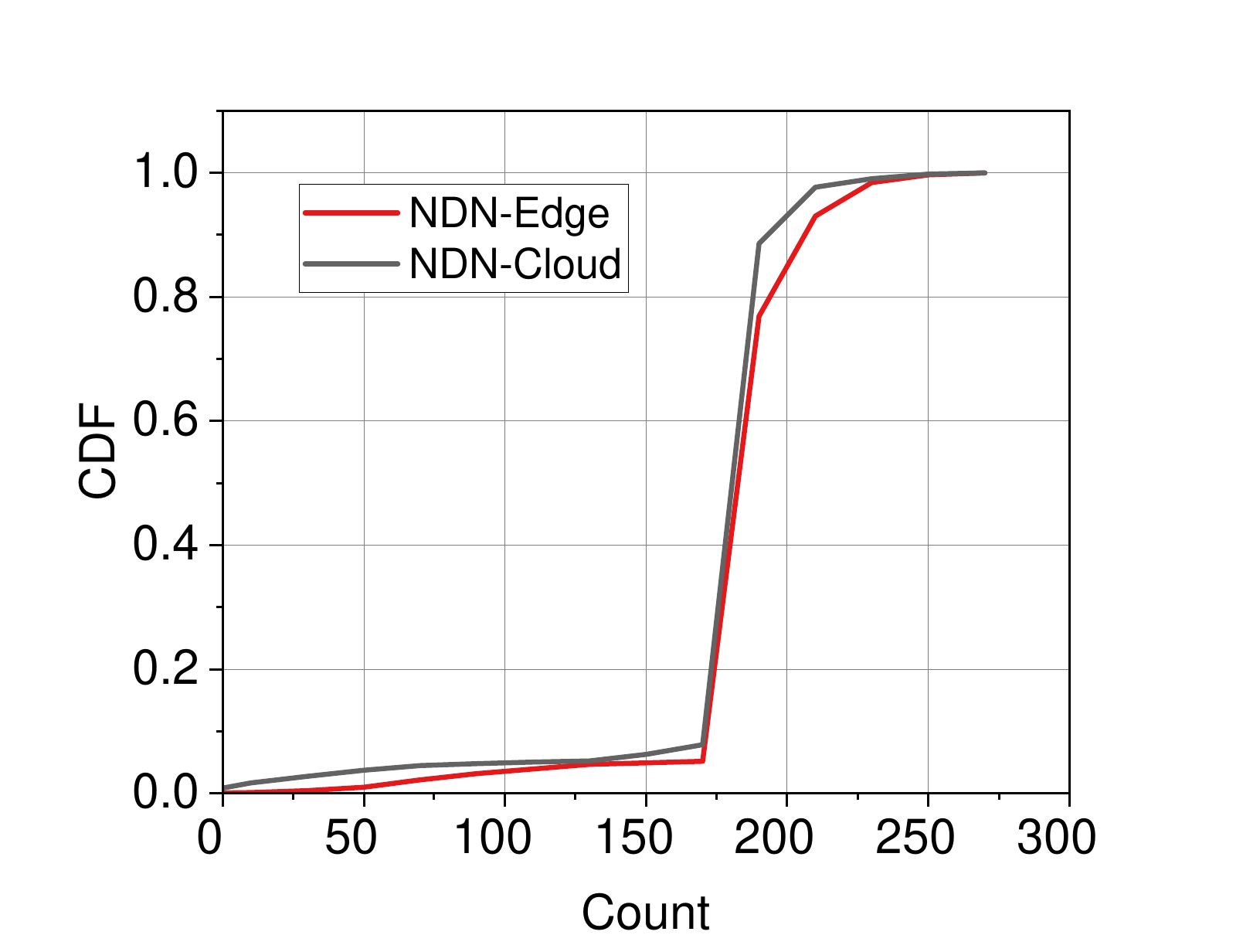}  
    \caption{Count of cache hits}
    \label{fig::cache-60rps}
\end{figure}

Figure~\ref{fig::cache-60rps} shows the cache hits of NDN-Edge and NDN-Cloud.
The P95 cache hits of NDN-Edge is 223 while that of NDN-IP is only 210.
This result indicates that distributed DT can further improve the cache hit.
Since NDN-Edge is close to the consumers, response latency is much faster and cache hit is improved.

%% file: conclusion.tex
\section{Conclusion}~\label{sec::con}
In this work, as an early-stage investigation, we examine the performance of using distributed DTs in NDN networks for data query.
Our insight is that cloud DTs require data to be routed and processed in a remote location, making data communication a bottleneck for DTs.
In some cases (e.g., healthcare), data cannot be shared due to privacy regulations and hence distributed DT is a good fit because data resides where it is produced.
Apart from that, NDN is also a good fit for DTs.
NDN focuses on data rather than location, and DTs need to exchange massive amounts of data between the physical and virtual entities continuously, making data-centric NDN more efficient than host-centric networking.

We implemented two scenarios over NDN and IP-based networks, respectively.
Our simulations demonstrate that NDN significantly enhances DT performance, reducing response latency by 10.2$\times$ over IP-based networks. Additionally, although each edge DT is less powerful than the cloud DTs (each edge DT can only serve one type of query),
it further cuts latency by 46.7\% compared to cloud-based DTs, highlighting the benefits of distributed architectures.

%% file: main.bbl
\begin{thebibliography}{10}

\bibitem{YSUJOI20}
Ibrar Yaqoob, Khaled Salah, Mueen Uddin, Raja Jayaraman, Mohammed Omar, and
  Muhammad Imran.
\newblock Blockchain for digital twins: Recent advances and future research
  challenges.
\newblock {\em IEEE Network}, 34(5):290--298, 2020.

\bibitem{MSCDD22}
Antonino Masaracchia, Vishal Sharma, Berk Canberk, Octavia~A. Dobre, and
  Trung~Q. Duong.
\newblock Digital twin for 6g: Taxonomy, research challenges, and the road
  ahead.
\newblock {\em IEEE Open Journal of the Communications Society}, 3:2137--2150,
  2022.

\bibitem{chen2024}
Chen Chen, Manuel Herrera, Ge~Zheng, Liqiao Xia, Zhengyang Ling, and Jiangtao
  Wang.
\newblock Cross-edge orchestration of serverless functions with probabilistic
  caching.
\newblock {\em IEEE Transactions on Services Computing}, 17(5):2139--2150,
  2024.

\bibitem{liang2023}
Hengshuo Liang, Cheng Qian, Chao Lu, Lauren Burgess, John Mulo, and Wei Yu.
\newblock Named data networking (ndn) for data collection of digital
  twins-based iot systems.
\newblock In {\em 2023 IEEE/ACIS 21st International Conference on Software
  Engineering Research, Management and Applications (SERA)}, pages 122--127,
  2023.

\bibitem{chen2023}
Chen Chen, Lars Nagel, Lin Cui, and Fung~Po Tso.
\newblock S-cache: Function caching for serverless edge computing.
\newblock EdgeSys '23, page 1–6, New York, NY, USA, 2023. Association for
  Computing Machinery.

\bibitem{GTEK24}
Qi~Guo, Fengxiao Tang, Tiago~Koketsu Rodrigues, and Nei Kato.
\newblock Five disruptive technologies in 6g to support digital twin networks.
\newblock {\em IEEE Wireless Communications}, 31(1):149--155, 2024.

\bibitem{awsdt}
AWS.
\newblock Aws iot twinmaker, 2025.

\bibitem{Azuredt}
Microsoft.
\newblock Azure digital twins, 1 2025.

\bibitem{RSK20}
Adil Rasheed, Omer San, and Trond Kvamsdal.
\newblock Digital twin: Values, challenges and enablers from a modeling
  perspective.
\newblock {\em IEEE Access}, 8:21980--22012, 2020.

\bibitem{ASMTKSAL19}
Rafael~Gomes Alves, Gilberto Souza, Rodrigo~Filev Maia, Anh Lan~Ho Tran, Carlos
  Kamienski, Juha-Pekka Soininen, Plinio~Thomaz Aquino, and Fabio Lima.
\newblock A digital twin for smart farming.
\newblock In {\em 2019 IEEE Global Humanitarian Technology Conference (GHTC)},
  pages 1--4, 2019.

\bibitem{AAGGB20}
P.~Angin, M.~Anisi, F.~G{\"o}ksel, C.~G{\"u}rsoy, and
  A.~B{\"u}y{\"u}kg{\"u}lc{\"u}.
\newblock Agrilora: a digital twin framework for smart agriculture.
\newblock {\em J. Wirel. Mob. Networks Ubiquitous Comput. Dependable Appl.},
  11(4), 2020.

\bibitem{FSCCD22}
Muhammad Fahim, Vishal Sharma, Tuan-Vu Cao, Berk Canberk, and Trung~Q. Duong.
\newblock Machine learning-based digital twin for predictive modeling in wind
  turbines.
\newblock {\em IEEE Access}, 10:14184--14194, 2022.

\bibitem{NK23}
Gholamreza Nasiri and Abdollah Kavousi-Fard.
\newblock A digital twin-based system to manage the energy hub and enhance the
  electrical grid resiliency.
\newblock {\em Machines}, 11(3), 2023.

\bibitem{ISA23}
Reda Issa, Mostafa S.Hamad, and Mostafa Abdel-Geliel.
\newblock Digital twin of wind turbine based on microsoft® azure iot platform.
\newblock In {\em 2023 IEEE Conference on Power Electronics and Renewable
  Energy (CPERE)}.

\bibitem{HVBZKVM23}
Josua Höfgen, Birgit Vogel-Heuser, Fandi Bi, Jingyun Zhao, André Kraft, Bernd
  Vojanec, and Timo Markert.
\newblock Architecture of a versatile digital twin with socket-based
  communication and azure dt.
\newblock In {\em 2023 IEEE 19th International Conference on Automation Science
  and Engineering (CASE)}, pages 1--8, 2023.

\bibitem{WTS23}
Erdan Wang, Pouria Tayebi, and Yeong-Tae Song.
\newblock Cloud-based digital twins’ storage in emergency healthcare.
\newblock {\em International Journal of Networked and Distributed Computing},
  11(2):75--87, 2023.

\bibitem{DSK24}
Yash Deshpande, Eni Sulkaj, and Wolfgang Kellerer.
\newblock Twinran: Twinning the 5g ran in azure cloud.
\newblock {\em arXiv preprint arXiv:2407.13340}, 2024.

\bibitem{AMRN23}
Marica Amadeo, Marco Martalò, Giuseppe Ruggeri, and Michele Nitti.
\newblock Enabling social digital twins in the 6g era with information centric
  networking.
\newblock {\em IEEE Communications Magazine}, 61(12):112--117, 2023.

\bibitem{Amadeo2024}
Marica Amadeo, Giuseppe Ruggeri, Claudio Marche, and Michele Nitti.
\newblock Service discovery and provisioning in social digital twin networks: a
  name-based approach.
\newblock In {\em 2024 20th International Conference on Wireless and Mobile
  Computing, Networking and Communications}.

\bibitem{topologyzoo}
Simon Knight, Hung~X. Nguyen, Nickolas Falkner, Rhys Bowden, and Matthew
  Roughan.
\newblock The internet topology zoo.
\newblock {\em IEEE Journal on Selected Areas in Communications},
  29(9):1765--1775, 2011.

\bibitem{ndnSIM}
ndnSIM.
\newblock Named-data networking (ndn) simulator, 2025.

\bibitem{Riley2010}
George~F. Riley and Thomas~R. Henderson.
\newblock {\em The ns-3 Network Simulator}, pages 15--34.
\newblock Springer Berlin Heidelberg, Berlin, Heidelberg, 2010.

\bibitem{Michael2023}
Judith Michael, Maike Schwammberger, and Andreas Wortmann.
\newblock Explaining cyberphysical system behavior with digital twins.
\newblock {\em IEEE Software}, 41(1):55--63, 2024.

\end{thebibliography}
